\documentclass[article]{IEEEtran}
\IEEEoverridecommandlockouts
\usepackage{subcaption}
\usepackage{latexsym}
\usepackage{fancyhdr}
\usepackage[normalem]{ulem}
\usepackage{comment}
\usepackage{amsfonts,amssymb,amsmath,bm}
\usepackage{hyperref}
\usepackage[export]{adjustbox}
\usepackage{array}
\usepackage{float}
\usepackage{tabularx}
\usepackage{makecell}
\usepackage{stackengine}

\usepackage[justification=centering,font=scriptsize]{caption}
 \DeclareGraphicsExtensions{.pdf,.jpeg,.png,.eps}
\usepackage{enumerate}
\usepackage{graphics}
\usepackage{graphicx} 
\usepackage{cite}
\usepackage{mathrsfs}
\usepackage{physics}
\usepackage{stfloats}
\usepackage{balance}
\usepackage{upgreek}

\usepackage{stackengine,scalerel}
 \usepackage{placeins}
\usepackage{cite}
\usepackage{amsmath,amssymb,amsfonts}
\usepackage{algorithmic}
\usepackage{graphicx}
\usepackage{textcomp}
\usepackage{xcolor}
\usepackage{comment}
\def\BibTeX{{\rm B\kern-.05em{\sc i\kern-.025em b}\kern-.08em
    T\kern-.1667em\lower.7ex\hbox{E}\kern-.125emX}}
\makeatletter 
\newcommand{\linebreakand}{%
  \end{@IEEEauthorhalign}
  \hfill\mbox{}\par
  \mbox{}\hfill\begin{@IEEEauthorhalign}
}
\makeatother 

\begin{document}
\title {From Connectivity to Multi-Orbit Intelligence: Space-Based Data Center Architectures for 6G and Beyond
}

\author{Shimaa~Naser,~\IEEEmembership{Member,~IEEE,} Maryam~Tariq, Raneem~Abdel-Rahim, De~Mi, ~\IEEEmembership{Senior Member,~IEEE,} Azzam Mourad, ~\IEEEmembership{Senior Member,~IEEE,} Hadi~Otrok, ~\IEEEmembership{Senior Member,~IEEE,}  Mahmoud Al-Qutayri, ~\IEEEmembership{Senior Member,~IEEE,} Ayman~Elnashar, and      Sami~Muhaidat,~\IEEEmembership{Senior Member,~IEEE}
\thanks{ S. Naser, M. Tariq, R. Abdel-Rahim, and S. Muhaidat are with the KU 6G Research Center, Department of Computer and Information Engineering, Khalifa University, Abu Dhabi, UAE. (emails: shimaa.naser; 100036622; 100067007; sami.muhaidat@ku.ac.ae)}
\thanks{D. Mi is with the College of Computing, Birmingham City University, Birmingham, B4 7XG, U.K. (email: de.mi@bcu.ac.uk) }
\thanks{A. Mourad and H. Otrok are with the Department of Computer Science, Khalifa University, Abu Dhabi, UAE. (emails: azzam.mourad; hadi.otrok@ku.ac.ae)}
\thanks{M. Al-Qutayri is with the Systems-on-Chip Laboratory, Department of Computer and Information Engineering, Khalifa University, Abu Dhabi, UAE. (email: mahmoud.alqutayri@ku.ac.ae)}
\thanks{A. Elnashar is with Technology Strategy, Architecture \& Innovation Division, e\&, UAE. (email: ayelnashar@eand.com)}
}
\maketitle
\begin{abstract}
Direct handset-to-satellite (DHTS) communication is emerging as a core capability of 6G non-terrestrial networks, enabling standard devices to directly access low Earth orbit (LEO) satellites. While LEO provides the physical access layer for DHTS, large-scale device connectivity introduces challenges in mobility management, interference control, spectrum efficiency, and constellation-wide coordination. Relay-only LEO architectures are insufficient to manage massive handset access under dynamic traffic and energy constraints. This article introduces a hierarchical architecture in which direct handset-to-LEO access is supported by multi-orbit space-based data centers (SBDCs) spanning LEO, medium Earth orbit (MEO), and geostationary Earth orbit (GEO). In this framework, LEO satellites handle radio access and real-time inference, while higher orbital layers provide regional aggregation, global orchestration, and compute-aware routing. By embedding distributed in-orbit computing, energy-aware scheduling, and AI-driven hierarchical control, the constellation evolves from a passive relay network into an intelligent multi-layer system capable of supporting large-scale DHTS services. We discuss key enabling technologies, envisioned multi-orbit integrated Earth-space compute architecture, and open research challenges in integrating multi-orbit computing, highlighting pathways toward scalable and resilient 6G DHTS networks.

\end{abstract}

 \section{ Introduction}
Currently, Earth observation, remote sensing, and autonomous systems, to mention a few, are generating large volumes of data. A single Earth observation satellite captures over 100 terabytes of imagery daily, and aggregate space-generated data is projected to reach 566 Exabytes over the next decade \cite{cepurnaite2024space}. Similarly, with the widespread deployment of 5G and increasing reliance on data-intensive applications, terrestrial data generation is experiencing a steep growth trajectory. Global mobile data traffic is forecast to exceed 350 Exabytes per month by 2030, more than a fourfold increase within a single decade\cite{knightfrank2025datacentres}. As a consequence, terrestrial data centers (DCs) are expanding continuously to accommodate growing demand for storage, processing, and real-time analytics. Nevertheless, this growth scales energy consumption, cooling requirements, and operational costs, which increasingly limit the scalability and sustainability of purely ground-based computing infrastructure.
\begin{figure}[t] 
    \centering
    \begin{subfigure}{\columnwidth}
        \centering        \includegraphics[width=\columnwidth]{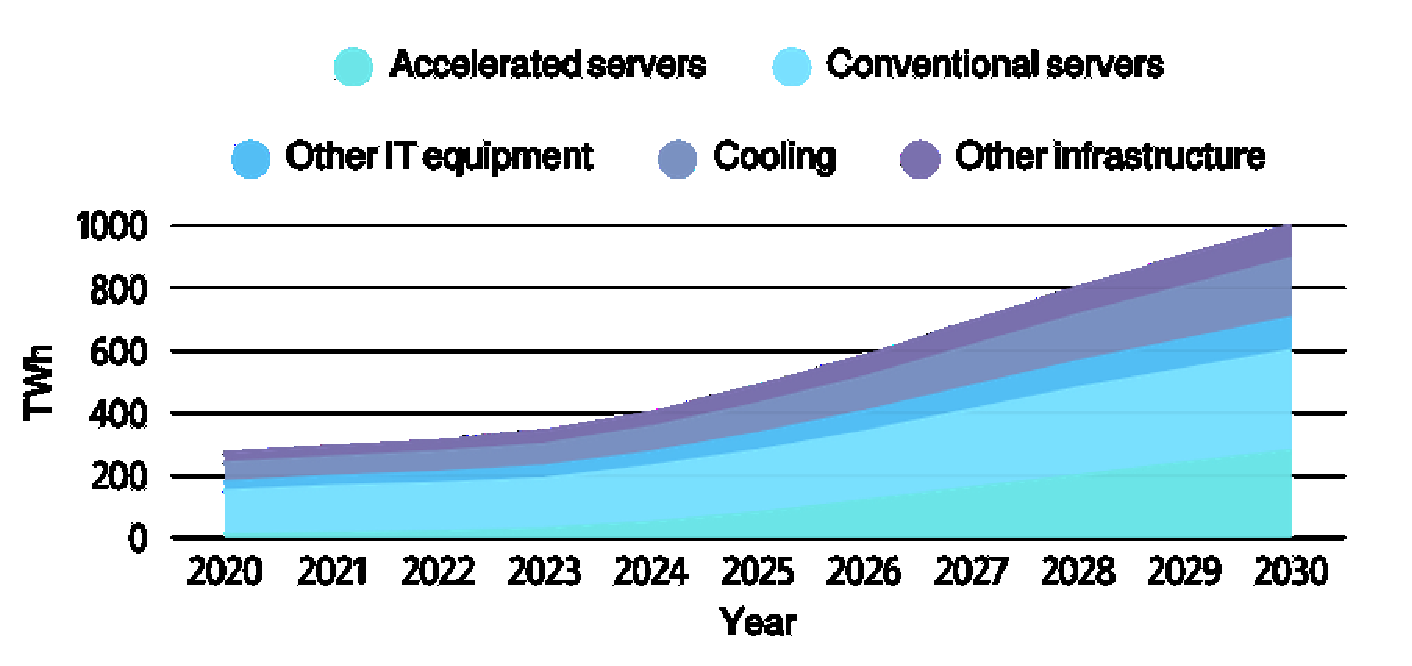} 
        \par (a) 
        \label{fig1a}
    \end{subfigure}

    \par\vspace{0.5em} 
    \begin{subfigure}{\columnwidth}
        \centering
        \includegraphics[width=\columnwidth]{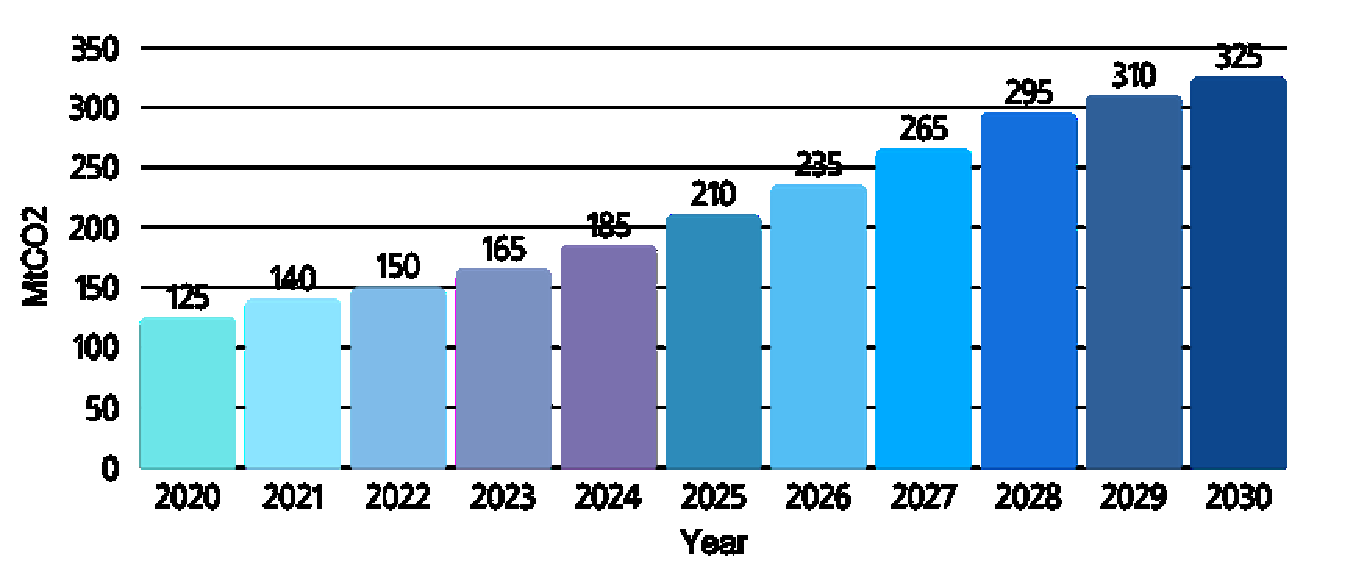} 
        \par (b)
        \label{fig1b}
    \end{subfigure}
    \caption{(a) Global data centers electricity consumption forecast 2020-2030. (b) Global data centers $\text{CO}_2$ emissions forecast 2020-2030 \cite{IEA2025EnergyAI}.  }
    \label{fig:example}
\end{figure}
\subsection{Terrestrial DCs Physical Limits}
The global electricity consumption of terrestrial DCs is forecast to nearly double from approximately 415 TWh in 2024 to over 945 TWh by 2030 (depicted in Fig. 1a), with AI-accelerated servers showing the steepest growth. Moreover, the additional deployment of terrestrial  DCs adds pressure on cooling infrastructures. In particular, air- and water-based cooling systems are becoming economically and environmentally costly, especially in regions experiencing water scarcity and rising ambient temperatures. The energy required for cooling further amplifies the total carbon footprint of terrestrial DCs. Although global efforts to enhance energy efficiency and integrate renewable energy solutions are ongoing, CO2 emissions from DCs are rising because absolute capacity growth outpaces incremental improvements, as illustrated in Fig. 1b.

It is also important to note that alternative terrestrial compute paradigms, edge and fog computing, alleviate latency for ground-based users, but do not resolve the fundamental energy, cooling, or coverage constraints, and are structurally bound to terrestrial infrastructure. Furthermore, they fall short in serving space-generated data in situ, and do not provide sovereign or jurisdiction-neutral storage. These limitations become more pronounced as non-terrestrial networks (NTN) expand to support large-scale direct handset-to-satellite (DHTS) access, where traffic demand and mobility management extend beyond the boundaries of terrestrial infrastructure. 
Table I compares three compute infrastructure paradigms across the dimensions most relevant to the next-generation NTN.
\begin{table*}[t]
\caption{Comparison of Compute Infrastructure Paradigms}
\centering
\renewcommand{\arraystretch}{1.2}
\begin{tabular}{|l|p{4cm}|p{4cm}|p{4.3cm}|}
\hline
\textbf{Dimension} & \textbf{Terrestrial DCs} & \textbf{Edge/Fog Computing (Earth)} & \textbf{Space-Based Data Centers (SBDCs)} \\ \hline

Primary Energy Source & Grid-dependent & Grid + localized renewables & Direct solar harvesting in orbit \\ \hline

Cooling Mechanism & Air/liquid cooling & Localized cooling & Radiative dissipation in vacuum \\ \hline

Scalability Constraints & Land availability, grid capacity, and zoning regulations & Urban density, infrastructure access, and deployment space & Launch capacity, orbital allocation, and space traffic management\\ \hline

Latency to Space Assets & High (via gateways) & High (Earth-bound) & Reduced via in-orbit processing \\ \hline

Latency to Remote Regions & Moderate--high & Reduced locally & Reduced through orbital coverage and in-orbit processing \\ \hline

Regulatory Scope & Governed by national energy, land-use, and environmental policies & Local/municipal operators & Spectrum, orbital, and cross-border compute governance \\ \hline

Role in NTN & Downstream processing & Terrestrial support & Integrated compute within NTN fabric \\ \hline

\end{tabular}
\end{table*}

\begin{table*}[b!]
\centering
\caption{Comparison of this work against prior SBDCs literature. }
\label{tab:comparison_SBDCs}
\renewcommand{\arraystretch}{1.2}
\begin{tabular}{lcccccc}
\hline
\textbf{Work} & \textbf{Multi-Orbit} & \textbf{AI Orchestration} & \textbf{Energy Model} & \textbf{Security} & \textbf{Regulation} & \textbf{Std. Alignment} \\ \hline
\hline
\cite{sym15071326} & x & x & \checkmark & x & x & x \\ \hline
\cite{Aili2025CarbonNeutralDataCentres}& x & x & \checkmark & x &  \checkmark  & x \\ \hline
\cite{Periola2019SpaceBasedDataCentres} & x & x & \checkmark & x & x & \checkmark \\ \hline
\cite{10642886} & \checkmark & x & x & x & x & x \\ \hline
\cite{9887918} & \checkmark & x & x & x & x & \checkmark \\ \hline
\textbf{This Paper} & \checkmark & \checkmark & \checkmark & \checkmark & \checkmark & \checkmark \\
\hline
\end{tabular}
\end{table*}

\subsection{NTN Compute Gap}
As terrestrial data infrastructure continues to scale, NTN has witnessed a profound transformation from traditional satellite communications to become an integral part of a broader architectural vision. In particular, DHTS communication is emerging as a foundational component of 6G NTN, enabling smartphones to directly access low-Earth orbit (LEO) constellations for ubiquitous connectivity. Nevertheless, existing NTN designs, as specified in  3rd Generation Partnership Project (3GPP) Releases 17–19 \cite{3gpp_ntn_overview,3gpp_edge_overview}, are still communication-centric: tasks such as image interpretation, traffic forecasting, network optimization, and artificial intelligence (AI) inference associated with LEO and GEO systems are offloaded to ground DCs for processing, albeit with some lightweight processing executed locally at the satellite. End-to-end application latency in LEO systems is inherently high, accumulating across feeder-link queuing, ground DC processing, and return transmission, which fundamentally limits real-time responsiveness. Furthermore, the tight power and bandwidth constraints mean that data volumes from modern Earth observation satellites far exceed Ka-band feeder link capacity within available visibility windows. In this context, the Phi-Sat-1 mission validates onboard inference as a direct mitigation, demonstrating meaningful downlink reduction through in-orbit data refinement.


Recently, industry and academia have been exploring the deployment of in-orbit space-based data centers (SBDCs) as a potential solution to these constraints. SBDCs drive down operational costs by using solar energy without the constraints of terrestrial solar farms. In addition, they can achieve efficient thermal dissipation through passive radiative cooling in space. Arguably most importantly, they can be scaled modularly and deployed widely without the physical and permitting challenges facing Earth-based infrastructure. Industry initiatives, including Amazon Project Kuiper with AWS integration, Google's Suncatcher project, Starcloud, the EU Ascend program, and the Three-Body Computing Constellation by Alibaba and Zhejiang Lab, reflect growing recognition of this potential. 

Beyond individual initiatives, the deeper architectural insight that differentiates SBDCs from both terrestrial DC expansion and conventional NTN is that SBDCs unify connectivity, compute, storage, and control within the NTN fabric itself. Merely scaling terrestrial DCs is insufficient because the data generation rate from space assets outpaces the downlink capacity, and jurisdictional constraints limit cross-border data flow. NTN alone is insufficient without in-orbit compute because it imposes the full terrestrial latency penalty and bandwidth bottleneck on every in-orbit service. In this regard,  SBDCs resolve both constraints simultaneously by embedding intelligence at the orbital layer where data is generated, transforming satellites from mere sensing and relay platforms into producers of active compute nodes. Rather than sending raw observation data, SBDCs can export inference results, learned feature representations, and model updates, thereby reducing bandwidth demand while delivering decision-ready information to terrestrial networks.

\subsection{Related Work and Contribution Positioning}
Previous research on SBDCs has examined feasibility from different angles, including cooling resource utilization \cite{sym15071326}, environmental sustainability and land-water footprint reduction \cite{Aili2025CarbonNeutralDataCentres, Periola2019SpaceBasedDataCentres}, architectural concepts for orbital edge and cloud computing \cite{Aili2025CarbonNeutralDataCentres,10642886}, and integration of DC functions within distributed satellite cluster networks \cite{9887918}. While these investigations show the promise of SBDCs, they focus mostly on isolated topics. Solving computational problems as a static or auxiliary service is addressed separately from network integration. Moreover, they are limited to single-orbit installations or rely heavily on terrestrial infrastructure, neglecting AI workload orchestration, multi-orbit interoperability, hardware aging, regulatory compliance, and standardization alignment.

The main contribution of this article is an integrated architectural framework that elevates SBDCs from standalone feasibility concepts to fully network-integrated computing entities. Specifically, this paper:
\begin{itemize}
\item Introduces a multi-orbit hierarchical computing architecture spanning LEO, MEO, GEO, and deep-space layers, with functional roles derived from orbital physics and system constraints.
\item Establishes a computation–communication–energy co-design perspective, showing how orbital dynamics and shared resource constraints reshape routing, scheduling, and workload orchestration relative to terrestrial cloud systems.
\item Provides representative SBDCs use cases, enabling technologies, and deployment models aligned with 6G NTN evolution.
\item Articulates open technical, operational, and regulatory challenges that must be addressed to transition SBDCs from conceptual prototypes to scalable, interoperable infrastructure.
\end{itemize}

\section{Integrated Earth–Space Computing Architecture}
The design of an SBDCs architecture differs fundamentally from conventional satellite systems and terrestrial clouds. Because computation, storage, and networking are co-located at orbital nodes and constrained by shared energy budgets and orbital dynamics, routing, scheduling, and energy management must be jointly optimized. The architecture in Fig. 2 is built around these coupled constraints.

\begin{figure}[t]
    \centering
    \includegraphics  [width=1\linewidth,height=10cm]{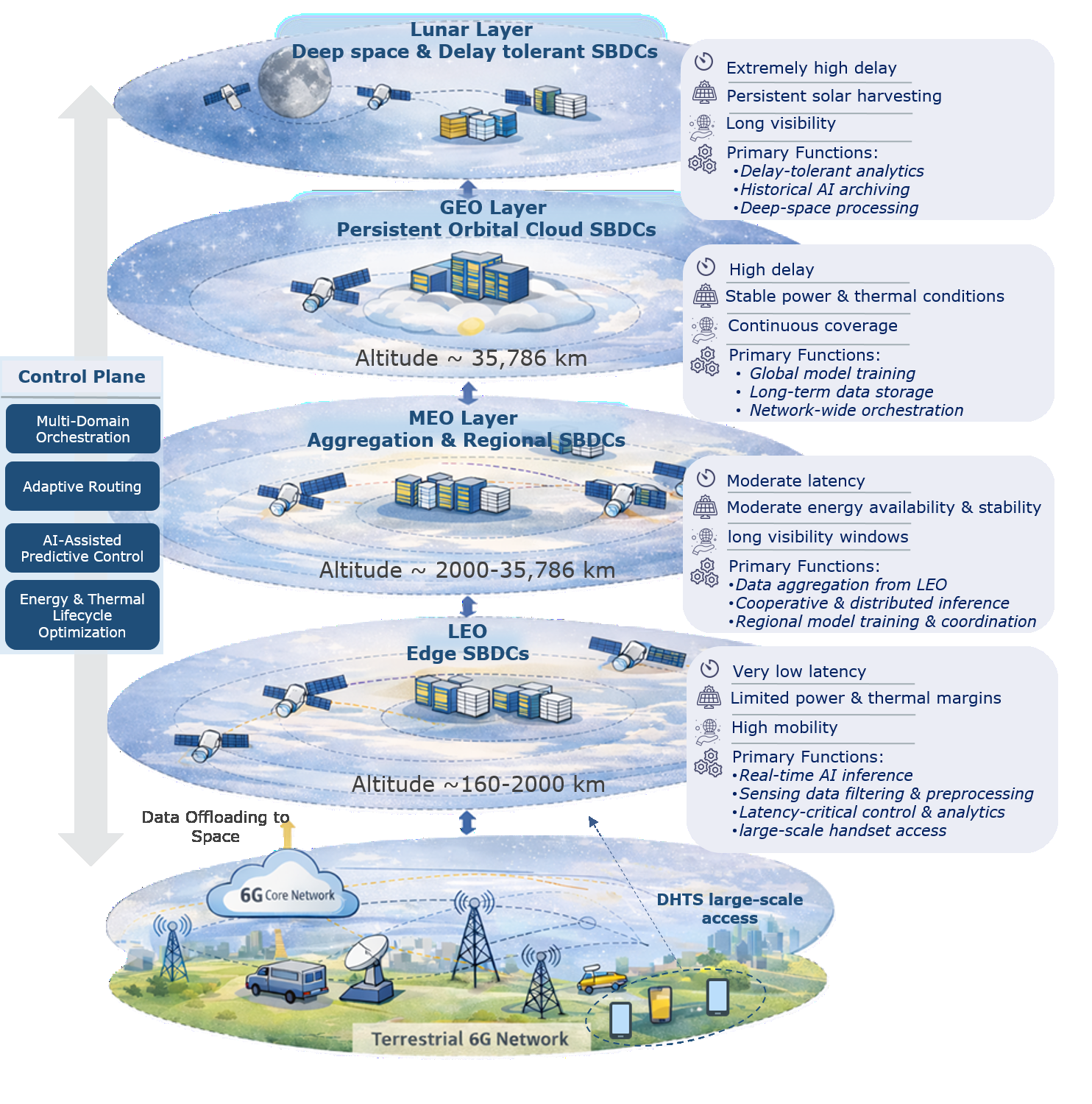}
    \caption{Multi-orbit Space-Based Data Center architecture supporting DHTS communication. }
    \label{fig:placeholder}
\end{figure}

\subsection{Orbital Layering and Functional Roles}
The assignment of SBDCs' functions to specific orbital layers is not an arbitrary architectural choice; it follows directly from the propagation physics, orbital mechanics, and energy availability that govern each regime. The central design principle is that each layer's physical characteristics define a distinct capability profile, and SBDCs' functions must be matched to layers whose profiles are consistent with the latency, continuity, and compute requirements.

The \textbf{LEO}  layer serves as the edge computing tier of the SBDCs architecture, providing low-latency sensing, direct handset access, and real-time inference. LEO nodes push computation close to the data source, including Earth observation sensors and DHTS mobile devices. With propagation delays of only a few milliseconds, LEO nodes host latency-critical functions such as real-time AI inference, integrated sensing and communication (ISAC), and fast control analytics. In DHTS scenarios, LEO satellites additionally act as the radio access layer, handling bursty uplink traffic, dynamic handovers, and interference coordination for large-scale mobile devices. However, the high orbital velocity limits ground visibility to short contact windows, during which application data, aggregated handset traffic, orchestration state, and telemetry must all be exchanged. Periodic eclipse phases further constrain energy availability; hence, LEO nodes are optimized for bursty, low-latency processing such as extracting compact features, generating inference outputs, and forwarding summarized results to higher orbital layers for further processing or long-term storage.


The \textbf{MEO} layer represents the fog computing tier of the SBDCs architecture, bridging the LEO edge and the GEO cloud. With ground latencies in the tens to low hundreds of milliseconds, longer visibility windows, and reduced eclipse exposure, MEO nodes function as persistent regional aggregators. With simultaneous visibility to multiple LEO nodes, MEO satellites aggregate geographically distributed inference outputs, forming a regional context that neither LEO nor GEO can efficiently establish alone. In DHTS operation, this extends to cooperative traffic processing across multiple LEO access satellites, enabling regional mobility coordination and load balancing. Stable MEO-to-MEO inter-satellite links (ISLs) further enable cooperative inference across regional clusters. MEO is therefore the layer where regional model refinement and coherence tasks are most effectively executed before consolidated outputs are forwarded to GEO for global orchestration.


The  \textbf{GEO} layer provides the global orchestration and long-horizon compute backbone. Fixed at geostationary altitude, GEO nodes deliver approximately a quarter second ground latency but offer persistent, hemisphere-wide coverage, near-zero eclipse fraction, and stable solar exposure. These properties make GEO the only tier capable of hosting continuous-availability workloads: long-term data storage with strict retention guarantees, global AI model training aggregating contributions from all orbital layers simultaneously, and network-wide orchestration that maintains a consistent view of the entire constellation state, including large-scale DHTS traffic patterns.
 \textbf{Lunar} and \textbf{cislunar} nodes extend the architecture to the deep-space regime where one-way propagation delays preclude real-time interaction. All computation at this layer must be delay-tolerant, and is best suited to archival storage, scientific data processing, and long-running simulations offloaded opportunistically when Earth visibility permits.

\subsection{Energy and Thermal State Abstraction}
In SBDCs, power is a harvested physical resource determined by orbital mechanics, including solar flux, panel orientation, and eclipse geometry, as well as battery state-of-charge, rather than operator decisions. This transition from power as a control variable to power as a state variable forces the compute load to adapt to energy supply; any orchestration system that ignores energy state will trigger either thermal runaway or underutilization.
Each node abstracts its energy status into a normalized operational zone. In the \textbf{\textit{green zone}}, adequate battery reserve and solar flux sustain any workload class, including compute-intensive AI inference. As the node approaches eclipse, it transitions to the \textbf{\textit{yellow zone}}, restricted to lightweight inference and interruptible compression tasks. In the \textbf{\textit{red zone}}  (deep eclipse or severely depleted reserves), the node executes only essential housekeeping and queues pending tasks for handoff via ISLs or local resumption after eclipse exit. Zone boundaries are dynamically adjusted based on predicted eclipse duration, time to next ground contact, and thermal state of the processing subsystem.
Thermal management is a fundamental constraint on compute capacity and tightly couples with energy management. Unlike terrestrial servers, orbital platforms dissipate heat only through radiation, governed by the Stefan–Boltzmann law, where rejection power scales as the fourth power of radiator temperature. These physical constraints are encoded as normalized thermal headroom descriptors alongside energy zone indicators and exposed to the orchestration layer as structured state, enabling resource-aware scheduling.
 \begin{figure}[ht]
    \centering
    \includegraphics[width=1\linewidth]{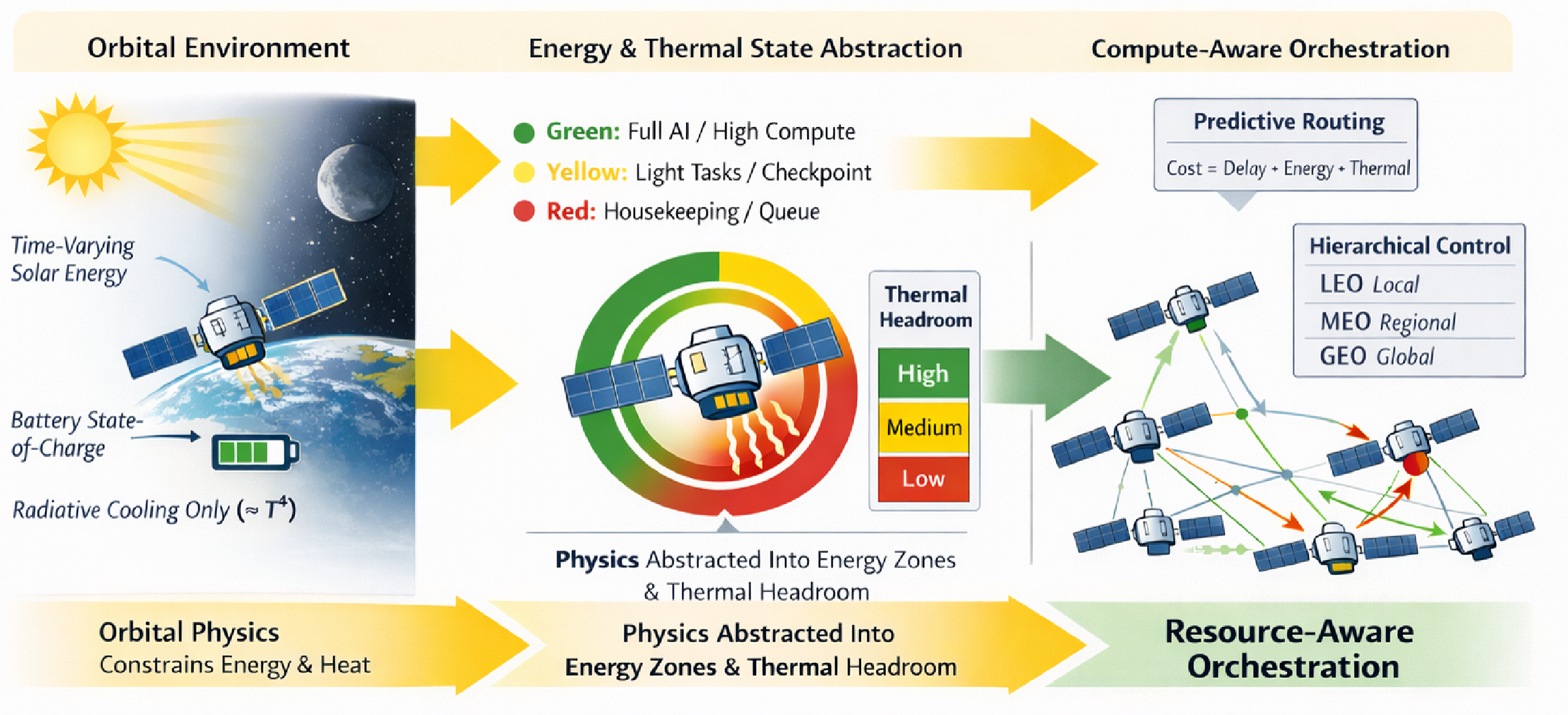}
    \caption{Resource-aware orchestration in SBDCs.}
    \label{fig:placeholder}
\end{figure}

\subsection{Compute-Aware Routing and Hierarchical Orchestration}

Conventional satellite routing paradigms, including virtual topology, virtual node, multilayer hierarchical protocols (e.g., MLSR, SGRP), and delay-tolerant networking (DTN)-based contact-graph routing \cite{s22124552}, optimize paths based solely on communication metrics such as propagation delay, link availability, queue occupancy, and hop count. This model is sufficient for relay-only constellations but breaks down in SBDCs, where node-level compute availability, energy zone, and hardware state directly affect service performance, particularly under large-scale DHTS traffic with frequent mobility-induced handovers. A path traversing a node in a depleted energy state or under heavy compute load may introduce indefinite queuing or task interruption, both effects invisible to link-layer metrics. Because orbital trajectories and contact windows are predictable, routing can be formulated over a time-expanded contact graph. In SBDCs, this graph must be augmented with compute-aware edge weights reflecting projected energy zones, available processing capacity, and thermal headroom. Routing then becomes a joint communication-and-computation placement problem: delay-sensitive flows follow low-latency ISLs paths, while compute-intensive tasks are steered toward nodes predicted to sustain execution, even if this requires longer communication paths.

This routing paradigm operates within a hierarchical control architecture aligned with orbital layering. LEO nodes execute forwarding and task decisions autonomously from precomputed tables during ground outages. MEO controllers coordinate regional routing and aggregation, maintaining constellation state across multiple LEO passes. GEO nodes maintain a global view, interface with terrestrial management entities (e.g., network data analytics function (NWDAF)), and enforce end-to-end policy and service-level agreement (SLA) compliance.

At the system level, orchestration becomes a stochastic, time-varying optimization over a joint compute-energy-communication state space. In principle, this can be expressed as minimizing a composite cost function over latency, energy consumption, and reliability risk, subject to thermal, bandwidth, and hardware constraints. Decisions include task placement, execution timing based on projected energy zones, replication for resilience, and checkpointing or migration triggered by degradation forecasts. Constraints span energy headroom, thermal limits, hardware health, ISLs bandwidth, and policy requirements. Exact global optimization is intractable at scale, which motivates structured approaches such as hierarchical control, model-predictive planning, and learning-based scheduling.

 \begin{figure*}
     \centering
     \includegraphics[width=0.9\linewidth]{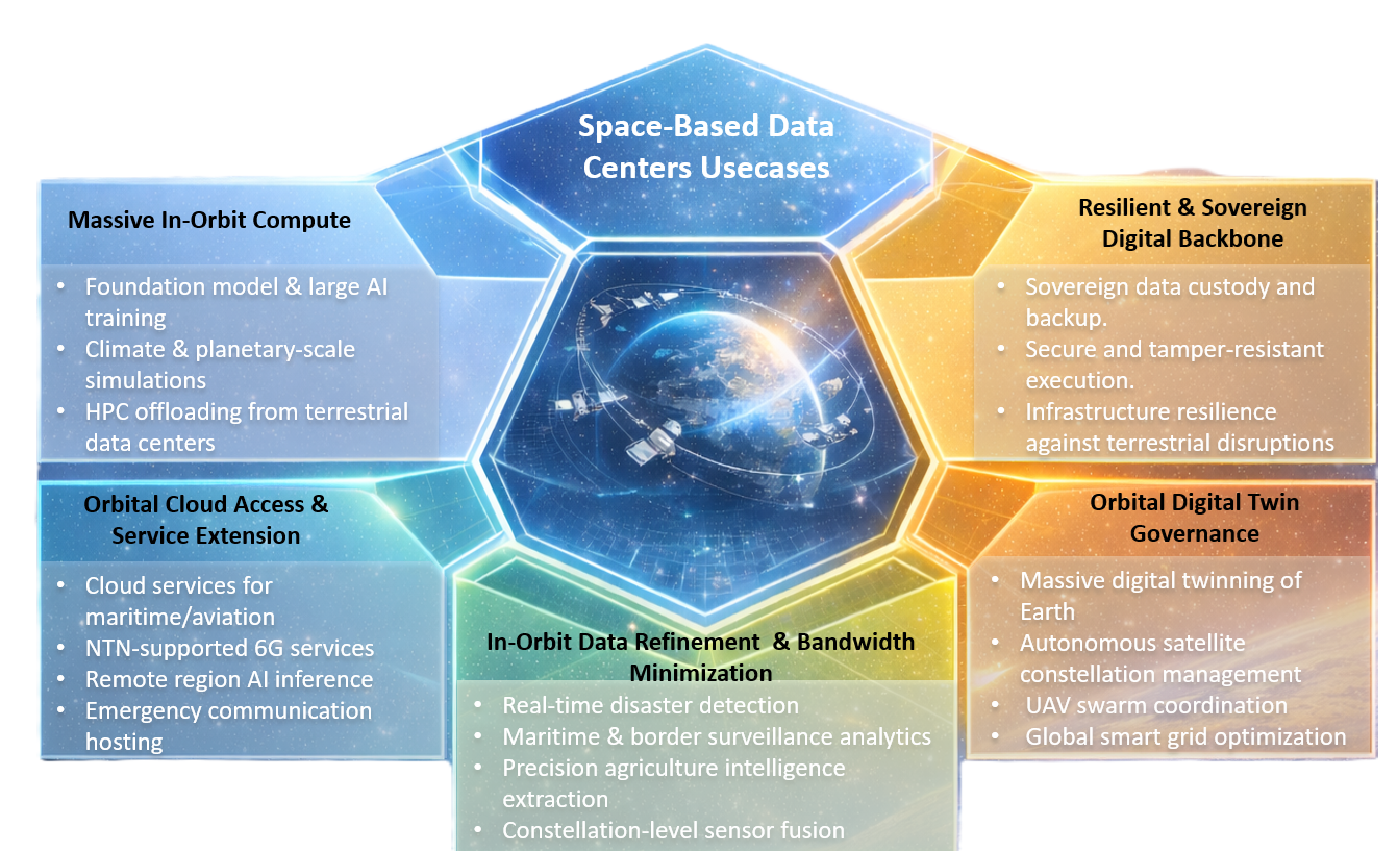}
     \caption{The envisioned emerging use cases for SBDCs. }
     \label{fig:usecases}
 \end{figure*}

\section{Envisioned  Use cases}
By offloading energy-intensive workloads to space and processing space-generated data in situ, SBDCs transform orbit into an active compute, storage, and intelligence layer, enabling capabilities beyond terrestrial DCs, as summarized in Fig.~\ref{fig:usecases}.

\begin{itemize}
    \item \textbf{In-Orbit Compute}
    
The energy-abundant orbital platforms allow hosting workloads that are unsustainable on Earth, transforming orbital space into a hyperscale computing tier. For instance, computationally intensive tasks such as large foundation model training, climate and Earth-system simulations, national-scale digital twins, and high-performance scientific computing can be offloaded via optical feeder links and NTN gateways to SBDC platforms for processing. Additionally, data generated from Earth observation and space weather monitoring can be processed in orbit, which reduces the burden of transmitting a large volume of raw data back to Earth.

\item  \textbf{Resilient \& Sovereign Digital Backbone}

SBDCs establish a physically isolated in-space repository protecting critical infrastructures, including government archives, banking ledgers, defense records, and healthcare data, which ensures service continuity during cyberattacks, geopolitical disruptions, or infrastructure collapse. Space data, including satellite telemetry and mission logs, can also be stored in SBDCs, reducing exposure to vulnerable ground facilities.

\item  \textbf{Orbital Cloud Access \& Service Extension
}

When terrestrial infrastructure is unavailable, congested, or disrupted, cloud services can be extended to orbit through SBDCs, transforming them into an orbital extension of the global network infrastructure. Leveraging NTN-supported 6G architectures, SBDCs provide broadband connectivity over oceans, airspace, polar regions, and remote areas where terrestrial networks are limited, enabling continuous cloud access for maritime navigation, aviation traffic control, and AI-assisted services. During natural disasters or geopolitical disruptions, SBDCs further host emergency communications and critical digital services to ensure operational continuity.


\item \textbf{In-Orbit Data Refinement \& Bandwidth Minimization}
Instead of transmitting large volumes of raw data to Earth, SBDCs process data directly in orbit through preprocessing, sensor fusion, anomaly detection, prioritization, and compression. Only key insights and alerts are offloaded to ground, while redundant data remains in space. This reduces feeder-link congestion, lowers backhaul demand, and shortens response times for delay-sensitive applications such as wildfire detection, flood monitoring, maritime surveillance, and border security. By sharing processed information across satellites, constellation-level analysis becomes possible, improving spectrum efficiency and easing the burden on terrestrial networks.

\item \textbf{ Orbital Digital Twin Governance}

SBDCs act as a supervisory intelligence layer maintaining synchronized digital replicas of large-scale terrestrial and orbital infrastructures, including smart grids, transportation corridors, UAV swarms, maritime fleets, and satellite constellations, which enables system-level modeling, stress testing, and near real-time optimization. Rather than executing isolated tasks, SBDCs evaluate global system behavior, anticipate cascading effects, and synthesize coordinated strategies across dispersed assets. This enables dynamic resource allocation, conflict mitigation, and mission-level optimization without relying on centralized terrestrial command centers.


\end{itemize}

\section{Enabling Technologies}
Integrating a set of new technologies across networking, communication, computing, and energy management enables the envisioned SBDC services and helps meet performance targets.
\subsection{High-Capacity Reliable Space Networking}
 Free-space optical (FSO) ISLs form the physical-layer backbone of the SBDCs mesh. In vacuum, loss is dominated by geometric beam spreading rather than atmospheric effects, making the link budget predictable via aperture size and transmit power. The main challenge lies in acquisition, pointing, and tracking (APT): over hundreds to thousands of kilometres, microradian errors cause complete beam miss, requiring fast steering mirrors and kHz closed-loop tracking. While constellation-scale APT has been demonstrated, reacquisition delays still cause stochastic outages, handled at the network layer through disruption-tolerant, store-and-forward routing over precomputed contact schedules. The time-expanded contact graph provides this framework, with APT performance defining edge reliability.

\textbf{\textit{Semantic}} and \textbf{\textit{task-oriented communications}} extend optimization to the application layer. Instead of prioritizing bit-perfect delivery, it transmits only the representation necessary for the receiver's task. For SBDCs, a node completing onboard inference sends alerts, feature vectors, or structured summaries, achieving potentially orders-of-magnitude reductions by operating at a higher representational layer. Realizing this requires joint design of inference models and transmission encoding, blurring the boundary between computation and communication.

\subsection{Space-Native Computing Platforms}
The core platform design trade-off is between radiation tolerance and compute performance. Full radiation hardening using redundant layouts, conservative margins, and specialized fabrication ensures reliability at the expense of reduced performance. While sufficient for housekeeping and attitude control, this is inadequate for AI inference or gradient aggregation. An emerging alternative is mitigation-based tolerance: using commercial or near-commercial silicon enhanced with error-correcting code (ECC) memory to correct bit flips, lock-step dual-core execution to detect logic errors, and FPGA configuration scrubbing to clear accumulated faults. This approach replaces a fixed performance ceiling with a tunable one where mitigation strength can be adjusted by workload, allowing higher residual error rates for background processing and stricter protection for safety-critical inference. 

Heterogeneous integration that combines CPU, GPU, FPGA, and neural accelerators addresses SBDC workload diversity. FPGAs handle signal preprocessing, GPUs accelerate matrix-heavy inference, and CPUs manage orchestration. A single-type platform forces inefficient mappings, whereas a heterogeneous platform aligns each task with its most efficient compute unit. Lightweight containerization enables dynamic workload placement, and radiation-adapted trusted execution environments (TEEs) provide multi-tenant isolation without full hypervisor overhead.


\subsection{Resilient and Secure Orbital Infrastructure}
In SBDCs, resilience and security must be intrinsic to the system design since radiation exposure, hardware aging, power variability, and dynamic connectivity make faults unavoidable in orbit. Continuous monitoring of computing, energy, and communication subsystems enables early anomaly detection, while predictive degradation models support proactive workload migration. Techniques such as checkpointing, selective replication, and adaptive task redistribution allow the system to tolerate transient errors and partial failures. Critically, because intermittent ground contact precludes human-in-the-loop intervention, static redundancy alone is insufficient. \textbf{\textit{Agentic AI}} addresses this gap by enabling orbital nodes to act as goal-directed autonomous agents that perceive system state, reason over fault scenarios, and execute multi-step recovery plans without ground intervention. This enables dynamic workload replanning, task handoff negotiation via ISLs, and energy allocation adaptation within a single pass. Resilience thus evolves from a static hardware property into an emergent behavioral capability driven by distributed autonomous intelligence embedded across the constellation.


Security is equally foundational. As SBDCs adopt cloud-like execution models and may involve multiple operators, they must guarantee hardware integrity, workload isolation, and protected communication across orbital and terrestrial domains. Secure boot mechanisms, hardware-rooted identities, TEEs, and remote attestation provide strong platform integrity. Moreover, end-to-end encryption across inter-satellite and feeder links protects data in transit, while \textbf{\textit{post-quantum cryptography} }and satellite-based \textbf{\textit{quantum key distribution}} offer long-term protection against emerging cryptographic threats.


\subsection{AI-Driven Orchestration and Digital-Twin Technologies}

AI orchestration and constellation digital twins are foundational enabling technologies for SBDCs, translating hierarchical control, multi-agent reinforcement learning, and model-predictive scheduling into operational capability. A constellation management digital twin maintains a continuously updated virtual replica of the SBDCs network that tracks node energy trajectories, thermal margins, hardware health, workload queues, and ISLs' availability using telemetry and orbital models. It supports safe policy training, predictive workload admission before orbital states are directly observable, and early anomaly detection through model–telemetry divergence. Maintaining synchronization under intermittent, bandwidth-limited ground contacts is a uniquely orbital systems challenge, and integration with 6G management-plane twins requires interoperable data models and APIs.

\textbf{\textit{Federated learning}} further enables distributed intelligence. As an orchestration mechanism, it trains placement, energy, and reliability models across nodes without centralizing raw telemetry. As a service, it enables distributed LEO training with hierarchical gradient aggregation over ISLs using compact updates rather than raw data transfer. This dual role tightly couples AI capability with orbital infrastructure and underpins in-orbit compute, orbital digital twin governance, and in-orbit data refinement.

\section{Challenges and Open Research Directions}
Integrating SBDCs into future NTN creates challenges beyond traditional satellite and cloud systems. 
\subsection{Technical Challenges and Directions}
\begin{itemize}

\item \textbf{ Orbital Environment Constraints}

Multi-year electronics 
must tolerate a total ionizing dose (TID) of 10--100~krad(Si), yet 
single-event upsets (SEUs) still cause continuous raw bit errors in 
multi-gigabyte accelerator memories within specification~\cite{jpl_radhard_handbook}. Radiators 
near 300\,K reject only 100--300\,W/m$^{2}$ \cite{gilmore_thermal_control}, so kW-class nodes 
require several square metres of radiator area, coupling compute 
density to spacecraft geometry. Additionally, orbital debris growth introduces collision risk that further reduces long-term node availability. These constraints motivate health-aware 
orchestration that uses ECC counters, thermal 
headroom, and aging indicators to drive checkpointing, migration, and 
replication across nodes with different radiation histories.

\item \textbf{Mobility-Induced Dynamic Changes}
Orbital layers introduce inherent latency–visibility trade-offs. While LEO nodes provide low propagation delay, their short contact windows limit the time available for task execution and results' delivery \cite{frontiers_starlink_params}. Consequently, tasks offloaded late in a visibility window may miss their return opportunity. Although Sec. II-C introduces contact-aware routing using time-expanded graphs, a \textit{key open challenge} is mobility-aware scheduling which predicts visibility, latency, and queue dynamics from orbital information and determines when and where tasks should be executed or migrated so that results meet application latency constraints.


\item \textbf{Security and Trust Constraints}

Sec. IV-C discussed platform-level protections such as secure boot, hardware-rooted identities, TEEs, and remote attestation. However, these mechanisms do not indicate which nodes remain trustworthy over time. Satellite lifetimes of 5–15 years means that compromised nodes may persist undetected, while radiation-induced faults and intermittent connectivity complicate monitoring and response. Distributed AI workloads further introduce threats such as model poisoning or gradient manipulation. A key \textit{research direction} is trust-aware orchestration, where node and path selection incorporates attestation history, fault indicators, and software-update status alongside latency and energy constraints.


\end{itemize}
\subsection{Operational Challenges and Directions}
\begin{itemize}

\item \textbf{Intermittent Ground Connection}

Because ground links are intermittent and delayed, SBDCs must 
execute fault detection, isolation, and recovery (FDIR) and 
workload management autonomously on board. Today, FDIR mechanisms rely on platform-specific, hard-coded thresholds that trigger safe modes when sensor readings exceed predefined limits. This reactive approach is too rigid for SBDCs constellations, where heterogeneous faults can propagate across nodes faster than ground operators can respond. A central research direction is to replace these ad hoc rule sets with 
principled, distributed control algorithms that coordinate FDIR, 
fail-safe policies, and health-management modules across many nodes, 
remaining stable under partial observability and delayed ground 
updates.

\item \textbf{Multi-Domain Operators}

Future SBDCs' deployments will span commercial LEO constellations, 
governmental GEO systems, and terrestrial cloud providers, each 
governed by different regulatory regimes, security classifications, 
and business models~\cite{espi_sbdc}. Coordinating across these 
domains requires compatible control planes and abstraction layers 
for infrastructures designed in isolation. Analogous to how the 
Border Gateway Protocol (BGP) lets independent Internet networks 
exchange routing policies without revealing internal topology, 
SBDCs require an inter-domain protocol through which operators 
advertise compute availability, trust posture, and SLA terms in a standardized, privacy-preserving form, 
enabling multi-operator schedulers to place and migrate workloads 
while respecting local policies and isolation requirements.

\end{itemize}

\subsection{Regulatory and Policy Challenges and Directions}
\begin{itemize}

\item \textbf{Joint Spectrum and Compute Regulation}

Current spectrum licensing frameworks allocate frequencies based on 
orbital position and transmission parameters (frequency, bandwidth, 
power, beam direction), agnostic to where computation occurs. SBDCs 
break this assumption: on-board processing can suppress, aggregate, 
or retime data streams before downlink, altering traffic volume, 
timing, and interference relative to a store-and-forward satellite 
under the same license, yet current frameworks offer no way to 
reflect this computational efficiency in spectrum assignment. A key 
direction is to model the coupling between workload placement and 
spectrum demand, quantifying how shifting filtering, compression, or 
inference from ground to orbit affects feeder-link capacity and 
interference, feeding these models into joint spectrum--compute 
planning tools for regulators and operators.

\item \textbf{Jurisdiction Constraints}

Data processed in orbit may originate in one jurisdiction, traverse 
ISLs registered under others, and be consumed 
elsewhere, yet existing regulations were written for fixed terrestrial 
servers, not dynamically migrating workloads~\cite{espi_sbdc}. An optimal placement in terms of latency or energy may therefore be legally 
impermissible. Compliance-aware orchestration addresses this by 
treating jurisdictional constraints as first-class scheduling inputs, 
encoded as machine-readable policy representations with data-locality 
rules, cross-border transfer restrictions, and permitted-operator 
sets, supported by tamper-evident provenance logging that provides an 
auditable chain of custody across the Earth--space pipeline.

\item \textbf{Standardization Gaps}

The 3GPP specifications for 
NTN and edge computing cover satellite 
connectivity and generic edge deployment~\cite{3gpp_ntn_overview,
3gpp_edge_overview}, but do not address how  SBDCs node advertises 
resources, accepts workload placement, or participates in joint 
communication--computation scheduling. A key direction is to develop 
SBDCs-aware abstractions within existing standards: resource 
representations augmented with orbital attributes, inter-domain 
interfaces for workload placement and handover, and consistent 
telemetry formats for latency, availability, and trust posture. Early 
coordination with 3GPP, the European Telecommunications Standards 
Institute (ETSI), and the Consultative Committee for Space Data 
Systems (CCSDS) is essential before proprietary mechanisms fragment 
the ecosystem.

\end{itemize}

\section{Conclusion}
This article presented a unified architectural vision that elevates SBDCs from isolated feasibility concepts to fully integrated platforms within future NTN, capable of supporting scalable DHTS communication. We introduced a layered Earth–space computing architecture, discussed enabling technologies, highlighted emerging use cases, and outlined key technical, operational, and regulatory challenges. 
By positioning SBDCs as an integral component of next-generation NTN, this work provides a system-level roadmap toward transforming space infrastructure into a scalable and sustainable extension of the global digital infrastructure.

\section{acknowledgment}
AI-assisted tools were used to improve the language and clarity of this manuscript.
\bibliographystyle{IEEEtran}
\bibliography{References}

\end{document}